\documentclass[12pt]{article}
\usepackage{times}
\usepackage{url}
\usepackage{scicite}
\usepackage{braket}
\usepackage[]{color}
\usepackage[]{graphicx}
\usepackage{subcaption}
\newcommand{\red}{}

\newenvironment{sciabstract}{%
\begin{quote} \bf}
{\end{quote}}

\newcounter{lastnote}
\newenvironment{scilastnote}{%
\setcounter{lastnote}{\value{enumiv}}%
\addtocounter{lastnote}{+1}%
\begin{list}%
{\arabic{lastnote}.}
{\setlength{\leftmargin}{.22in}}
{\setlength{\labelsep}{.5em}}}
{\end{list}}

\title{Stripe order in the underdoped region of the two-dimensional Hubbard model}

\author{Bo-Xiao Zheng$^{1,2\ast\dag}$, Chia-Min Chung$^{3\dag}$, Philippe Corboz$^{4\dag}$, Georg Ehlers$^{5\dag}$,\\
  Ming-Pu Qin$^{6\dag}$, Reinhard M. Noack$^{5}$, Hao Shi$^{6\dag}$, Steven R. White$^{3}$, \\
  Shiwei Zhang$^{6}$, Garnet Kin-Lic Chan$^{1\ast}$\\
\\
\normalsize{$^{1}$Division of Chemistry and Chemical Engineering, California Institute of Technology,}\\
\normalsize{Pasadena, CA 91125, USA}\\
\normalsize{$^{2}$Department of Chemistry, Princeton University, Princeton, NJ 08544, USA}\\
\normalsize{$^{3}$Department of Physics and Astronomy, University of California, Irvine, CA}\\
\normalsize{$^{4}$Institute for Theoretical Physics and Delta Institute for Theoretical Physics, University }\\
\normalsize{of Amsterdam, Science Park 904, 1098 XH Amsterdam, The Netherlands}\\
\normalsize{$^{5}$Fachbereich Physik, Philipps-Universit\"at Marburg, 35032 Marburg, Germany}\\
\normalsize{$^{6}$Department of Physics, The College of William and Mary, Williamsburg, VA 23187, USA}\\
\\
\normalsize{$^\ast$To whom correspondence should be addressed; E-mail:  boxiao.zheng@gmail.com}\\
\normalsize{$^\ast$To whom correspondence should be addressed; E-mail:  gkc1000@gmail.com}\\
\normalsize{$^\dag$These authors contributed equally to the calculations in this work.}
}

\date{}

\begin{document}
\baselineskip24pt
\maketitle

\begin{sciabstract}
Competing inhomogeneous orders are a central feature of correlated electron materials including the high-temperature superconductors.
The two-dimensional Hubbard model  serves as the canonical microscopic physical model for such systems.
Multiple orders have been proposed in the underdoped part of the phase diagram, which corresponds
to a regime of maximum numerical difficulty.
By combining the latest numerical methods in exhaustive simulations, we uncover the ordering
in the underdoped ground state. We find a stripe order that has a highly compressible wavelength
on an energy scale of a few Kelvin, with wavelength fluctuations coupled to pairing order. 
The favored filled stripe order is different from that seen in real materials.
Our results demonstrate the power of 
modern numerical methods to solve microscopic models even in challenging settings.
\end{sciabstract}

%\section*{Introduction}
Competing inhomogeneous orders are a common feature in many strongly correlated materials~\cite{Dagotto2005}.
A famous example is found in the underdoped region of the phase diagram of the
high-temperature cuprate superconductors (HTSC).  Here, multiple probes,
including neutron scattering, scanning tunneling microscopy, resonant X-ray
scattering, and nuclear magnetic resonance spectroscopy all lend support to various proposed inhomogeneous orders, such as
charge, spin, and pair density waves, with suggested patterns ranging from
unidirectional stripes to checkerboards~\cite{comin2016resonant,julien2015}. Recent
experiments on cuprates  indicate that the observed inhomogeneous
orders are distinct from, and compete with, pseudogap
physics~\cite{parker2010fluctuating,gerber2015three}.

%% Neutron scattering experiments first
%% indicated the existence of stripe type inhomogeneities in the 1/8 doping region of the cuprates in LSCO. In this early work, it was
%% not entirely clear whether or not the inhomogeneities were due to material imperfections
%% or a basic feature of the physics. In more recent years, improved crystals and additional experimental evidence such
%% as scanning tunneling microscopy and RIXS have further supported the presence of various kinds of inhomogeneous orders, including bond order waves,
%% checkerboard patterns, charge- and spin-density waves and stripes across all examples of cuprates. Thus while different patterns
%% appear in different cuprates, the presence of some kind of inhomogeneity appears to be incontrovertibly established.
%% In the cuprates in particular, recent evidence suggests that the inhomogeneity is driven by ground-state physics and can thus be
%% distinguished from the pseudogap physics at finite temperature. 

Much theoretical effort has been directed to explaining the origin of the
inhomogeneities~\cite{Fradkin2015}.  Numerical calculations on
microscopic lattice models have provided illuminating examples of the
possible orders.  The prototypical lattice model to understand HTSC is
the 2D Hubbard model on a square lattice, with the Hamiltonian
\begin{equation}
  H = -\sum_{\langle ij \rangle,\sigma \in \{\uparrow,\downarrow\}} t
  a^\dag_{i\sigma} a_{j\sigma} + U \sum_i n_{i\uparrow}
  n_{i\downarrow}
\end{equation}
where $a^{\dag}$ ($a$)  denote the usual fermion creation (annihilation) operators, $n$ is the number operator, and $t$ and $U$ are the kinetic
and repulsion energies. A large number of numerical techniques have
been applied to compute the low-temperature and ground-state phase
diagram of this model.
Early evidence for unidirectional stripe ordering in the Hubbard model
came from Hartree-Fock calculations~\cite{Poilblanc1989,Zaanen1989,machida89,schulz89},
whereas non-convex energy versus filling curves  in exact
diagonalization of small clusters of the $t$-$J$ model (derived from the Hubbard model at large $U$ where
double occupancy is eliminated)
were interpreted as signs of macroscopic phase separation~\cite{Emery1990,Emery1990a}.  Since then, inhomogeneous
orders have been obtained both in the Hubbard and $t$-$J$ models from calculations using the
density matrix renormalization group (DMRG)~\cite{white1998density,white2003stripes,hager2005stripe},
%variational and projector quantum Monte Carlo
variational quantum Monte Carlo \cite{himeda02} and constrained path auxiliary field
quantum Monte Carlo (AFQMC) \cite{chang2010spin},
infinite projected entangled pair states
(iPEPS)~\cite{Corboz2014}, density matrix embedding theory
(DMET)~\cite{zheng2016}, and functional renormalization
group~\cite{yamase2016coexistence} among others, although the type of
inhomogeneity can vary depending on the model and numerical method.  %% these studies, as
%% in some HTSC's, 1/8 doping represents a point of maximal inhomogeneity, as well as maximum numerical difficulty, in many 
%% of the simulations. 
However, there are other sophisticated simulations, for example, with
variational and projector quantum Monte Carlo~\cite{Sorella2002,hu12}, and
cluster dynamical mean-field theory, which do not see, or are unable to resolve, the inhomogeneous
order~\cite{macridin2006phase,LeBlanc2015}.  The most recent studies with
iPEPS~\cite{Corboz2014} and DMET~\cite{zheng2016}, as well as some
earlier variational calculations~\cite{himeda02,raczkowski07,chou08,chou10}, further show that both homogeneous
and inhomogeneous states can be stabilized within the
same numerical methodology, with a small energy difference between
homogeneous and inhomogeneous states, on the order of $\sim 0.01t$ per
site.

%% Subsequent works on the Hubbard model with DMRG~\cite{Jeckelmann}, variational and projector quantum Monte Carlo
%% (QMC)~\cite{chang2010spin}, projected entangled pair states
%% (PEPS)~\cite{Corboz2014}, and density matrix embedding theory (DMET)
%% calculations have found the presence of stripes, 
%% although with somewhat differing results for the inhomogeneity patterns, e.g. in stripe filling, or site-centered 
%% versus bond-centered stripes.
%% %% these studies, as is also found
%% %% in some HTSC's, 1/8 doping represents a point of maximal inhomogeneity, as well as maximum numerical difficulty, in many 
%% %% of the simulations. 
%% However, there are other sophisticated simulations, for example, with variational
%% and projector quantum Monte Carlo~\cite{Sorella2002}, and cluster dynamical
%% mean-field theory, which fail to see inhomogeneous states~\cite{LeBlanc2015}.
%% The most recent studies with iPEPS~\cite{Corboz2014} and DMET~\cite{Zheng2016} 
%% further show that both homogeneous and inhomogeneous states can be observed and
%% stabilized within the same numerical methodology, with a small energy
%% difference between homogeneous and inhomogeneous states, on the order of 
%% $\sim 0.01t$ per site. 

%% As any computation necessarily contains approximations, it has been difficult to decide whether or 
%% not the observed inhomogeneities are artifacts of the approximations, or true features of the ground-state phase diagram. 
The small energy differences between orders means that very small biases in ground state simulations, such as from an incomplete
treatment of fluctuations, using insufficiently accurate constraints to control the
sign problem, or from finite size effects, 
can easily stabilize one order over the other. Similarly, the low temperatures needed to resolve between orders is a challenge for
 finite temperature methods~\cite{white1989numerical,wu2016controlling}.
Settling the resulting debate between candidate states  has thus so far been beyond reach.
%% For example,
%% early quantum Monte Carlo methods were prevented from
%% studying the low temperature regime by the fermion sign problem[ref S.R. White, D.J.~Scalapino, R.L.~Sugar, E.Y.~Loh, 
%%     J.E.~Gubernatis, and R.T.~Scalettar,  ``Numerical study of the two-dimensional Hubbard model,'' 
%% {\sl Phys. Rev.} B{\bf 40}, 506 (1989).]
In this work %using advances in computational algorithms and computer processing power, 
we demonstrate  that, with the latest numerical techniques, obtaining a
definitive characterization of the ground state order in the underdoped
region of the 2D Hubbard model is now an achievable goal. As a representative
point in the phase diagram, we choose the well-known $1/8$ doping point at strong
coupling ($U/t=8$). Experimentally, this doping corresponds to a region of maximal
inhomogeneity in many HTSC's, and in the strong coupling regime it is
recognized as a point of maximum numerical difficulty and uncertainty in
simulations~\cite{LeBlanc2015}.  
Using state-of-the-art computations with
detailed cross-checks and validation, including newer methodologies such as infinite projected-entangled pair states
(iPEPS) and density matrix embedding theory (DMET) as well as recent developments in
established methodologies such as constrained-path auxiliary field quantum Monte Carlo
(AFQMC) and density matrix renormalization group (DMRG), and with exhaustive accounting for finite size effects combined
with calculations directly in the thermodynamic limit, we are able
to %% achieve unprecedented accuracy in this challenging region of the
%% ground-state phase diagram. In so doing, we can
finally answer the question: what is the order and physics found in the underdoped ground state of the 2D
Hubbard model? 

\section*{Computational strategy}

An important strategy we bring to bear on this part of
the Hubbard model phase diagram is to combine the insights of multiple numerical tools with complementary
strengths and weaknesses. This approach, pioneered in \cite{LeBlanc2015},
greatly increases the confidence of the numerical characterization. To understand what each method
contributes, we briefly summarize the theoretical background and corresponding sources of error. 
Further details are provided in \cite{supplementary}.

\noindent {\bf Auxiliary field quantum Monte Carlo}. AFQMC expresses the ground state of a finite system 
through  imaginary time evolution, $\lim_{\beta \rightarrow\infty} e^{-\beta H} | \Phi_0\rangle$, where $|\Phi_0\rangle$ is an initial state.
The projection is Trotterized, and the evolution reduces to a stochastic single-particle evolution in the presence of 
auxiliary fields generated by the Hubbard-Stratonovich decoupling of the Hubbard repulsion. 
Away from half-filling, this decoupling has a sign problem. 
We use the constrained path (CP) approximation to eliminate the sign problem at the cost of a bias 
dependent on the quality of the trial state~\cite{Zhang1995,chang2008prb}. 
In this work, the Trotter error is well converged and we report the statistical error bar.
%% To assess the finite size error 
%% we use different boundary conditions on the cluster, and 
To minimize the constrained path bias, we use several different trial states, including  self-consistent optimization of the trial state~\cite{qin2016coupling}. 
The calculations are carried out on finite cylinders with open, periodic, and twist-averaged boundary conditions, 
with widths of up to 12 sites, and lengths of up to 72 sites.
This method can reach large sizes and finite size effects are minimized. The uncontrolled
error is from the CP approximation. 
%The uncontrolled error sources are the finite size error and 
%the constrained path bias.

\noindent {\bf Density matrix renormalization group}. DMRG is a variational wavefunction approximation using
 matrix product states (MPS), which are low-entanglement states with a 1D entanglement structure. The
quality of the approximation is determined by the bond dimension (matrix dimension) of the MPS.
The calculations are carried out on finite cylinders with widths of up to 7 sites, and lengths of up to 64 sites, 
with periodic boundary conditions in the short direction and open boundaries 
in the long direction. Two different DMRG algorithms were used: one working in a pure (real-space) lattice basis, and
another in a mixed momentum/lattice (hybrid) basis, with the momentum representation 
used along the short periodic direction~\cite{Motruk2016}.
We remove  the bond dimension error and finite size error
in the long direction by well-known extrapolation procedures, and report the associated error bar~\cite{stoudenmire2012studying}. 
Consistency between the lattice and hybrid DMRG algorithms provides a strong validation of this error bar.
The remaining uncontrolled error is the finite width error in the periodic direction.

\noindent {\bf Density matrix embedding}. DMET is a quantum embedding method which 
works directly at the thermodynamic limit, although interactions are only accurately treated within 
an impurity cluster~\cite{knizia2012density}. To solve the impurity problem, consisting of 
a supercell of the original lattice coupled to a set of auxiliary bath sites, we
use a DMRG solver. We treat supercells with up to 18 sites. The error bar reported in DMET 
corresponds to the estimated error from incomplete self-consistency of the impurity problem. 
The remaining uncontrolled error is the finite impurity size error.

%% which helps in the measurement of order parameters. . Calculations on the embedded impurity cluster size
%% have a significantly smaller error than on the bare cluster, however, the cluster shape can nonetheless bias the calculations to
%% a given phase. This bias can be removed by extrapolating in the cluster size.

\noindent {\bf Infinite projected entangled pair states}. 
iPEPS is a
variational approach using a low-entanglement tensor network ansatz
natural to 2D
systems~\cite{verstraete2004,nishio2004,jordan2008classical}.  The
calculations are carried out directly in the thermodynamic limit where
different supercell sizes including up to 16 sites are used to stabilize
different low-energy states (with different orders commensurate with the supercell). As in DMRG, the accuracy of the ansatz is
systematically controlled by the bond dimension $D$ of the
tensors. Estimates of quantities in the exact $D$ limit are obtained
using an empirical extrapolation technique which is a potential source
of uncontrolled error.

%% iPEPS is a variational wavefunction approximation using
%% a  low-entanglement ansatz natural to 2D systems~\cite{jordan2008classical}. The calculations are carried out directly in the thermodynamic limit,
%% but to accelerate convergence we also use a simulation supercell of up to 25 sites.
%% The main error in iPEPS comes from the finite bond dimension, which, similarly to in DMRG, can be extrapolated
%% away while introducing an uncertainty; other errors (e.g. contraction error) are here controlled to
%% high precision. We report the extrapolation error bar; note that there are no sources of uncontrolled error.

% it is possible 
%% thus the only error in iPEPS comes from the finite bond dimension in the calculation. For finite
%% bond dimensions, the unit cell used in the iPEPS can bias the phase towards a given order. Extrapolation can either
%% performed in the bond dimension of the unit cell size to remove this error, or larger unit cells can be used. 

%% The primary biases are listed in \red{Table ~\ref{tab:bias}}.
\noindent {\bf Cross-checks: systematic errors, finite size biases}. The use of multiple techniques allows
us to estimate the uncontrolled errors from one technique using information from another. For example,
by carrying out simulations on the same finite clusters in the AFQMC and DMRG calculations, we can estimate
the constrained path bias in AFQMC. Similarly, in the AFQMC calculations we can treat larger width cylinders
than in the DMRG simulations; thus we can estimate the finite width error in DMRG.

In all of the methods, there is a bias towards orders commensurate with the shape of the simulation cell, be 
it the finite lattice and boundary conditions in AFQMC/DMRG, or the impurity cluster in DMET,  or the supercell in iPEPS.
Using this bias, together with different boundary conditions and pinning fields, we
can stabilize different meta-stable orders. For example, by setting up clusters commensurate 
with multiple inhomogeneous orders and observing the order that survives, we can determine the relative energetics of 
the candidate states. We can fit the orders along the short axis or the long axis of the cluster to obtain 
two independent estimates of the energy. We have carried out exhaustive studies of about 100
different combinations of clusters, cells, and boundary conditions, 
to fully investigate the low-energy landscape of states. 
These detailed results are presented in \cite{supplementary}.
To characterize the orders, we use the local hole density $1 - (\langle n_{\uparrow}+n_\downarrow\rangle)$,  
magnetic moment $\frac{1}{2} \langle n_{\uparrow}- n_{\downarrow}\rangle$,  and pairing order 
$\frac{1}{\sqrt{2}}(a_{i\uparrow}^{\dagger}a_{j\downarrow}^{\dagger}+a_{j\uparrow}^{\dagger}a_{i\downarrow}^{\dagger})$
%$\frac{1}{\sqrt{2}}\sum_{\sigma} \langle a^\dag_{i\sigma} a^\dag_{j\bar\sigma}\rangle$
($i$ adjacent to $j$).

%% The energies, order parameters, and calculation details
%% are reported in the supplementary information.
%% \blue{In the case of the AFQMC and DMRG calculations, which do not break particle number symmetry naturally, we perform seperate calcualtions with
%%   pinning pairing field on the open boundaries to estimate the existense of pairing order (supplementary information). \textit{This could be, but wasn't done in AFQMC}.}
%In the case of the AFQMC and DMRG calculations, which do not break particle number symmetry,
%we estimate the pairing order from the pairing correlation function (supplementary information).

\section*{Characterizing the ground state at 1/8 doping}

\begin{figure}[htpb]
  \centering
  \includegraphics[width=\columnwidth]{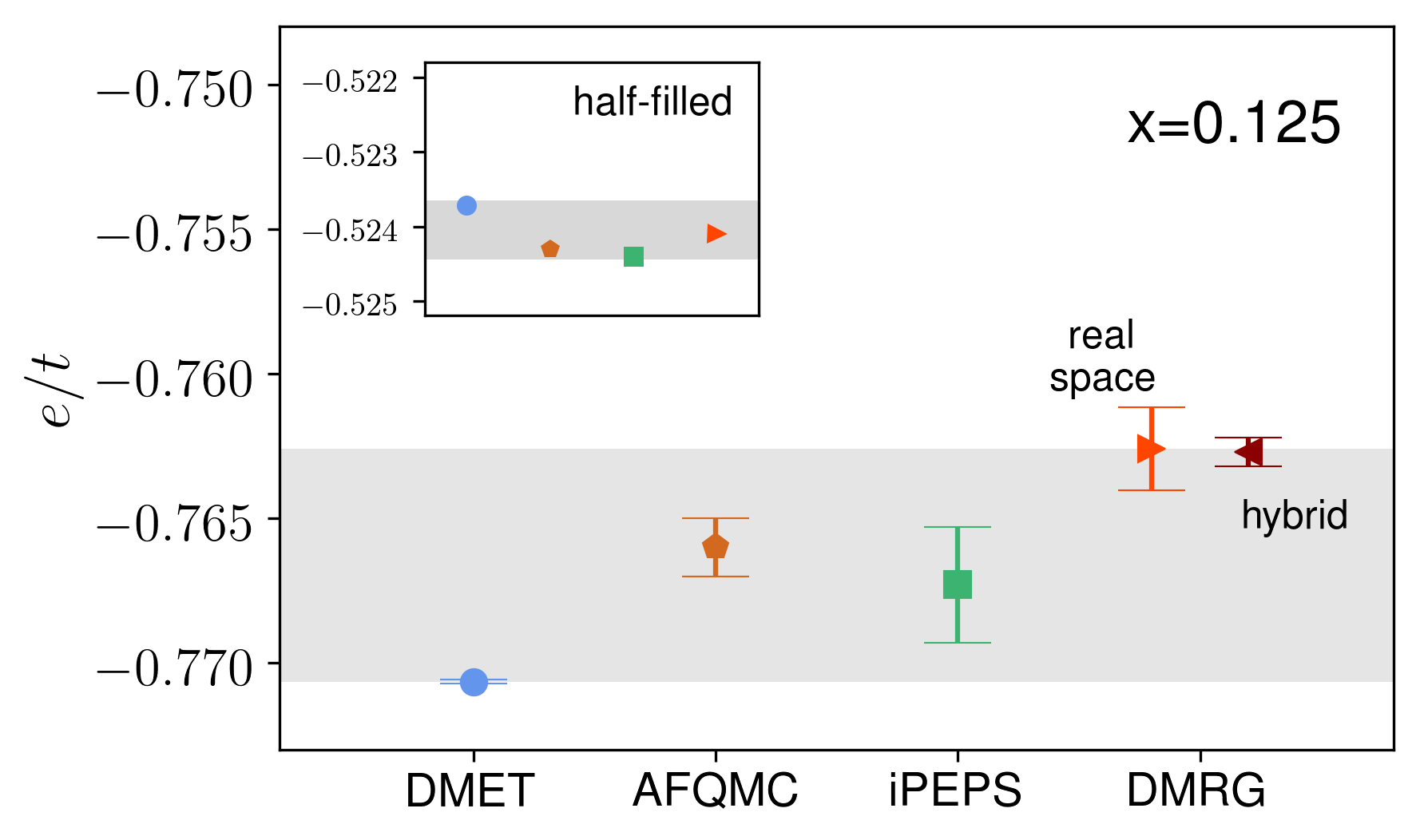}
  \caption{{\bf Ground state energies.} Best estimates of ground state energy for the $1/8$-doped 2D Hubbard model at $U/t=8$ from DMET, AFQMC, iPEPS and DMRG in units of $t$. 
    %Also shown (open symbol) is the AFQMC corrected for the constrained-path bias.
    Inset: Best estimates of ground state energy
  for the half-filled 2D Hubbard model at $U/t=8$.
\red{Here and elsewhere, error bars indicate only the estimable numerical errors of each method; uncontrolled systematic errors are not included.} For details see \cite{supplementary}.
}
  \label{fig:energy_fig}
\end{figure}

Using the above methods, we carried out calculations for the ground state of the 2D Hubbard model at 1/8 doping at $U/t=8$.
The first check of reliability is the independent convergence of the methods for the energy per site.
Although the quality of the ground-state energy may be a poor proxy for the quality of the corresponding state when the overall
accuracy is low (as there are always many degenerate states far above the ground state), calculations
with well-converged energies 
tightly constrain the ground state order, as
any degeneracies must be below the energy convergence threshold.
Figure~\ref{fig:energy_fig} shows the best energy estimate for the ground state from
the different methods~\cite{supplementary}. The two different DMRG formulations (real-space and hybrid basis)
are in good agreement, providing a strong independent check of the calculations: in subsequent figures
we report only the single consistent result.
Note that the error bars for AFQMC, DMRG, and DMET do not
reflect the uncontrolled systematic errors in the methods. However,
as described above, the systematic errors can be estimated by cross-checks between the methods.
For example, DMRG and AFQMC calculations on finite clusters with identical boundary conditions
provide an estimate of the small constrained path bias (see \cite{supplementary} and Ref.~\cite{qin2016coupling}) consistent
with the difference in the DMRG and AFQMC energies in Fig.~\ref{fig:energy_fig}; similarly
AFQMC extrapolations to the thermodynamic limit indicate that the DMRG energies are essentially
converged with respect to cylinder width.

%% For example, the difference between AFQMC and DMRG is consistent with the constrained phase bias
%% in simulations on finite cells with 
%%  (see supplementary materials and Ref. \cite{qin2016coupling}). This suggests that the small residual
%% difference is likely from the CP bias, and that the width-6 cylinders treated in DMRG are essentially converged
%% for the energy.

%thus we also show the AFQMC energy corrected for the constrained path bias from DMRG; the finite-width corrected
%DMRG calculation (from infinite width AFQMC extrapolations) is indistinguishable from the DMRG result on the plot.
%% with error bars denoting some estimated uncertainties for the methods. \blue{For CP-AFQMC and DMET, the error bars
%% does not reflect the the finite size error; for DMRG, we report the lowest extrapolated (with respect to cylinder length) energy;
%% the iPEPS energy, on the other hand does not have a finite size error.}
%Note that the CP-AFQMC, DMRG, and DMET
%error bars do not include an estimate of the finite size error; we only report the lowest energies observed in the calculations
%on finite clusters and impurities; the iPEPS energy on the other hand does not have a finite size error.

There is good agreement between all the methods, and all energies lie in the range 
$-0.767 \pm 0.004t$. If, for a typical HTSC material, we estimate $t \sim 3000K$, then this corresponds to a range
of about $\pm 10K$ per site, or $\pm 100K$ per hole. For a numerical comparison, this is also more than an order of magnitude lower than the   temperatures accessible in finite temperature, thermodynamic limit simulations in this part of the phase diagram, indicating
  that we are potentially accessing different physics~\cite{LeBlanc2015,wu2016controlling}.
Shown in the inset are the corresponding best estimates at half-filling
from the same methods, where the spread in energies is less than $0.001t$. This illustrates the significantly greater numerical
challenge encountered in the underdoped region. 
Nonetheless, the accuracy and agreement reached here represents a ten-fold improvement over 
recent comparisons of numerical methods at this point in the phase diagram~\cite{LeBlanc2015}.

%% \red{Nonetheless, the close agreement between all the methods, to within the estimated error bars, indicate that the sources of error
%% are well accounted for. The consensus (average) energy is [X]. Shown in the inset is the corresponding energy
%% estimates at half-filling, where the deviation between different methods is about [X], illustrating
%% the significantly greater difficulty of obtaining accurate results in the }. \blue{\textit{Actually the agreement is not very good. If we use finite size DMRG energy (-0.7552(8)), the agreement is better.}} 

\begin{figure}[htpb]
  \centering
  %\begin{subfigure}[t]{0.6\columnwidth}
  %  \includegraphics[width=\textwidth]{relative}
  %  \caption{Relative energy.}
  %\end{subfigure}
  %\begin{subfigure}[t]{0.3\columnwidth}
  %  \includegraphics[width=\textwidth]{DMET_sc}
  %  \caption{DMET uniform d-wave state.}
  %\end{subfigure}
  \includegraphics[width=\columnwidth]{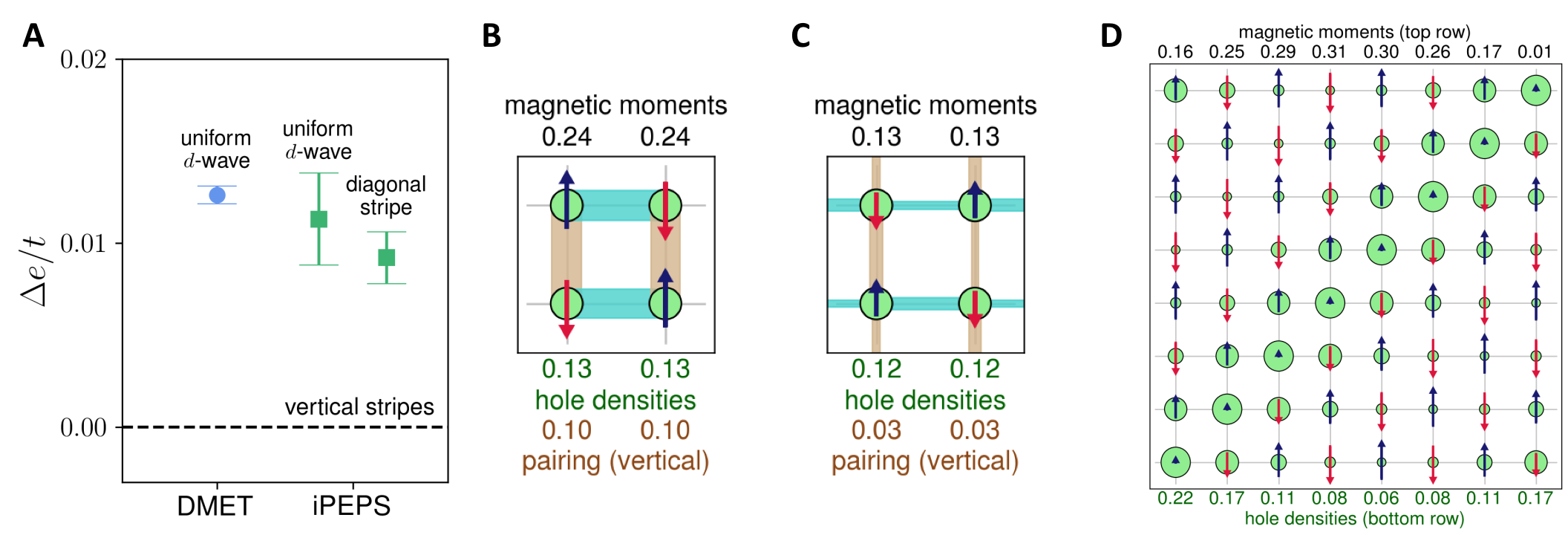}
  \caption{{\bf Competing states.} Shown are the energies of important competing states relative to the striped ground state from DMET and iPEPS and the sketches of the corresponding orders.
	  {\bf(A)} Relative energy of competing states in units of $t$ compared to the vertical striped state.
	  Charge, spin and pairing orders of the uniform $d$-wave state from {\bf(B)} DMET (blue circle) and {\bf(C)} iPEPS (green squares).
	  {\bf(D)} Charge and spin orders of the diagonal striped state from iPEPS. Note that
  the spins are flipped in the neighboring supercells. (For {\bf B}, {\bf C}, {\bf D}, circle radius is proportional to hole density, arrow height is proportional to spin density, bond width is proportional to pairing density). For more details see \cite{supplementary}.}
  \label{fig:relative_energies}
\end{figure}

%% \blue{Due different finite size errors, the absolute energies of the competing states are not generally comparable. Instead, 
%%   we illustrate the low energy states from each method with similar boundary conditions in Fig.~\ref{fig:low_e}. The relative orders of these states provide more information about the
%% competition. }

%% \begin{figure}[htpb]
%%   \centering
%%   \includegraphics[width=\columnwidth]{low_e_states}
%%   \caption{Low energy states for $1/8$-doped 2D Hubbard model at $U/t=8$ from DMET, iPEPS, DMRG (real space and hybrid space) and CP-AFQMC.
%%     In the plot, $w-\lambda$ denotes stripe state with wavelength $\lambda$. The DMET and iPEPS series come
%%   from $\lambda\times2$ impurity or simulation cell calculations. One exception is the unifrom d-wave state from DMET, which is extracted from $2\times2$ impurity calculation.
%% The CP-AFQMC series is computed with $2\lambda\times8$ finite lattice using twisted average boundary condition (TABC). The real space and hybrid DMRG calculations are performed
%% on width-6 cylinders, with the height extrapolated to infinity.}
%%   \label{fig:low_e}
%% \end{figure}

 \noindent {\bf Ground state stripe order}. For all the methods employed, 
 the lowest energies shown in Fig.~\ref{fig:energy_fig} correspond to a vertical striped state.
This is a co-directional charge and spin-density wave state, with the region of maximum hole density 
coinciding with a domain wall in the antiferromagnetism. 
%% . To within
%% the estimable error bars, there is near degeneracy  wavelengths between $5$ and $10$.
%% ; and
%% to within statistical or extrapolation errors in the methods, the stripe is a
%%  wavelength 8 (w8) vertical stripe (charge period 8, spin period 16).
%% (In the DMRG calculations, the w8 stripe is degenerate with the wavelength 7 (w7) stripe to within  
%% $0.0004t$, less than half of the DMRG extrapolation uncertainty, while
%% in CP-AFQMC, 
%% and can be seen to be very similar between the different techniques.
As mentioned, unidirectional stripes of various kinds are a long-standing candidate order
in the doped Hubbard and related models. Hartree-Fock calculations give filled stripes (i.e. one hole per charge unit cell)
in both vertical and diagonal orientations, whereas one of the first applications of the DMRG to 2D systems found strong evidence for half-filled stripes in the $t$-$J$ model~\cite{white1998density}.
Finally, one of the earliest examples of inhomogeneity in doped HTSC's were the
static half-filled stripes in LaSrCuO at 1/8 doping~\cite{tranquada1995evidence}.

%% Unidirectional striped orders have some history as a candidate order in microscopic models.
%% Early evidence for unidirectional filled stripes (i.e. one hole per charge unit cell) in the Hubbard model, 
%% orientations, came from Hartree-Fock calculations~\cite{Poilblanc1989,Zaanen1989}, while 
%% in  Early arguments for phase separation in the $t$-$J$ model~\cite{Emery1990,Emery1990a} can also be seen
%% as consistent with stripe formation once finite size effects are taken into account.

The convergence to the same inhomogeneous order in the ground state in the current study, from multiple
methods with very different approximations, strongly suggests that stripes indeed represent the true ground state order of
the Hubbard model in the underdoped regime, and further highlights the accuracy we achieve with different techniques.  
 However, the stripe order we find has some unusual characteristics. 
We return to the details of the stripe order, its associated physics, and its relationship with experimentally observed
stripes further below. First, however, we examine the possibility of other competing meta-stable states.

\noindent {\bf Competing states: uniform $d$-wave state}. %The first question to address is whether the ground-state order is homogeneous or inhomogeneous.
Recent work using iPEPS and DMET on the $t$-$J$ and Hubbard models suggested close competition between a 
 uniform $d$-wave superconducting ground state and a striped order~\cite{Corboz2014, zheng2016}.
 Uniform states did not spontaneously appear in any of our calculations which indicates that they
lie higher in energy than the striped order. We found that we could stabilize a uniform $d$-wave state
in the DMET calculations by constraining the impurity cluster to a $2\times 2$ or $2\sqrt{2} \times \sqrt{2}$ geometry
and in the iPEPS calculations by using a $2\times2$ unit cell. DMET calculations on similarly
shaped larger clusters (such as a $4\times 4$ cluster) spontaneously broke symmetry to create a non-uniform state. From
these calculations we estimate that the uniform state lies $\sim 0.01t$ above the lowest energy state, and is not competitive
at the energy resolution we can now achieve~\cite{supplementary}.

\noindent {\bf Competing states: other short-range orders}.  Although other types of order have been proposed in the underdoped region,
such as spiral magnetic phases~\cite{Chubukov1995,yamase2016coexistence} and checkerboard order~\cite{Vojta2002},
we find no evidence for other kinds of short-range orders at this point in the phase diagram. The lack  of 
checkerboard order, which would easily fit within the large clusters
in our simulations (e.g. up to $64 \times 6$ in the DMRG calculations)
appears to rule them out as low energy states, in agreement with earlier DMRG simulations on the $t$-$J$ model~\cite{white2004checkerboard}.
Though we cannot rule out incommensurate orders, we have found that the variation of energy with unit cell wavelength (see below) is 
quite smooth, thus we do not expect a dramatic energy gain from incommensurability. We note that studies that have found
incommensurate magnetic orders have focused on smaller values of $U$~\cite{yamase2016coexistence}.

\noindent {\bf Diagonal versus vertical stripes}. We find the ground state order to be a vertical stripe type order, but other
studies of stripes indicate that different orientations can form~\cite{Kato1990}. 
On short length scales, the relevant question is whether diagonal stripes (with a $(\pi, \pi)$ wave
vector) are competitive with vertical stripes (with a $(0, \pi)$ wavevector).
With the boundary conditions used in this work, diagonal stripes would be frustrated in the DMRG and AFQMC calculations,
and did not spontaneously appear.
To stabilize diagonal stripes in the DMET and iPEPS calculations, we used tilted $n\sqrt 2 \times \sqrt 2$  impurity clusters
($n=2,5)$ for DMET, and a
$16 \times 16$ simulation cell with 16 independent tensors in
iPEPS.
The $16 \times 16$ iPEPS cell
gave a diagonal stripe (Fig.~\ref{fig:relative_energies}) that was
significantly higher in energy than the vertical stripe, by $0.009t$.
The DMET cluster gave rise to a frustrated diagonal order that we also estimate to be
higher in energy by $\sim 0.005t$~\cite{supplementary}.
%% $5\sqrt 2 \times \sqrt 2$ DMET cluster and $5 \times 5$ iPEPS
%% cell both  although the precise order was slightly different (see supplementary information).
%% %% Comparing the energy of tilted and non-tilted clusters in DMET is complicated by the different finite cluster errors.
%% We estimate that the diagonal stripe state is higher in energy by $0.005t$ (DMET) -- (iPEPS).
Although it is likely that the orientation of the stripe will depend on doping and coupling, 
vertical stripes appear to be significantly preferred at this point in the phase diagram.

\begin{figure}[htpb]
  \centering
  \includegraphics[width=\columnwidth]{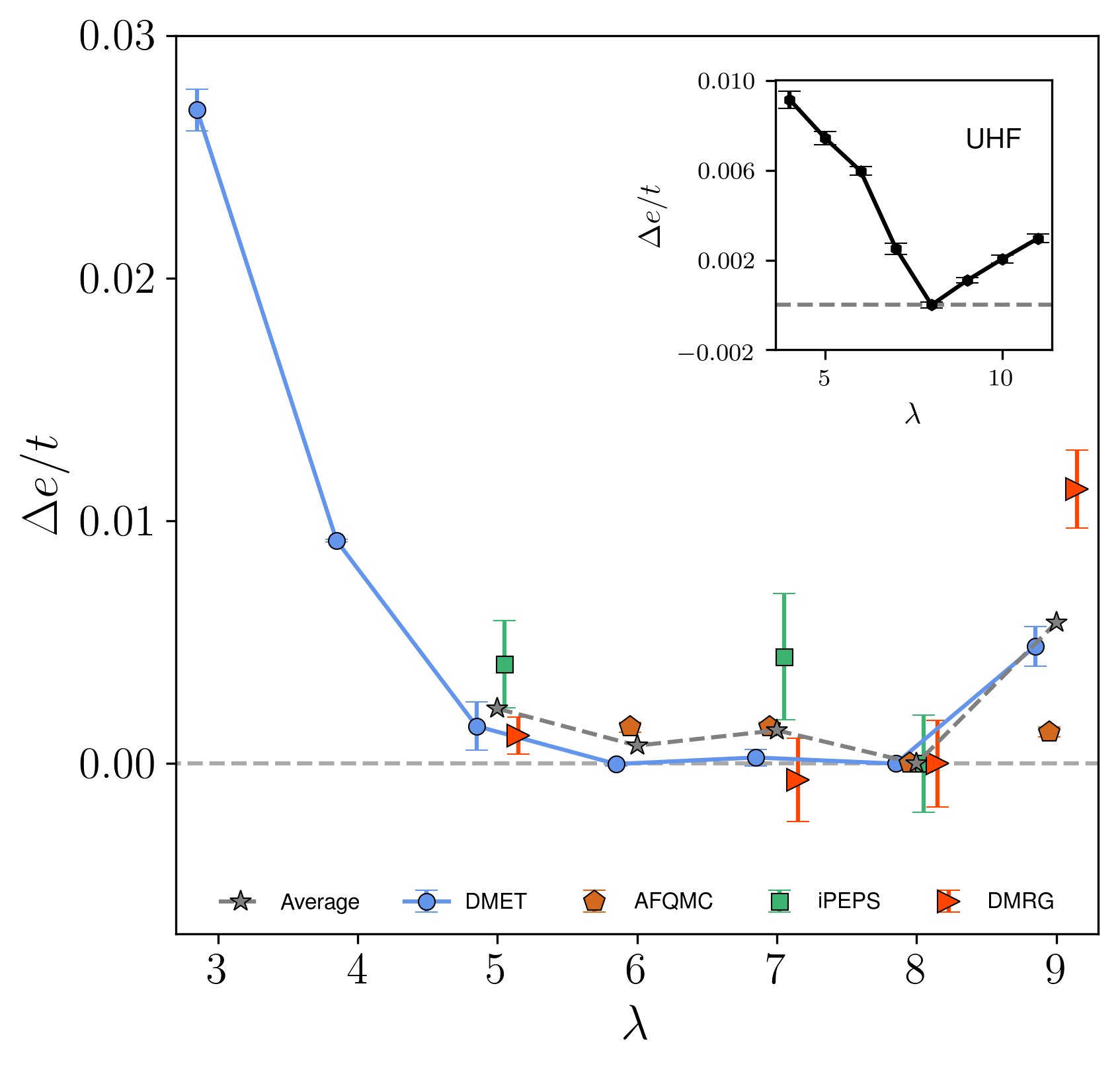}
  \caption{{\bf Wavelength of the vertical stripe order.} Energies of stripes with different wavelengths relative to that of the wavelength 8 stripe from DMET, AFQMC, iPEPS and DMRG in units of $t$. To aid readability, the data points are shifted horizontally. Inset: Relative energies of stripes with different wavelengths from UHF, with an effective coupling $U/t=2.7$. For details of calculations, see \cite{supplementary}.}
  \label{fig:energy_wavelength}
\end{figure}

\noindent {\bf Ground state stripes: detailed analysis}. We now return to a more detailed discussion of the
vertical stripe order found in the ground-state. Within the family of vertical stripes, 
we can consider questions of wavelength (charge and spin periodicity), type and strength of charge and spin modulation
(e.g. bond- versus site-centered), and coexistence with pairing order.

We first discuss the wavelength $\lambda$. At $1/8$ doping, the filling of the stripe is related to the wavelength by $\lambda/8$. %Although the lowest energy state appears to be a wavelength 8 stripe, 
As described, we can access different wavelength meta-stable stripes and their relative energetics by 
carefully choosing different total cluster dimensions and boundary conditions (in the DMRG and AFQMC calculations) or unit cell/impurity sizes (in the iPEPS and DMET calculations)~\cite{supplementary}. 
%% The DMET unit cells could
%% stabilize stripes with wavelengths between 3-9, while the iPEPS unit cells could stabilize stripes of wavelength 5, 7, 8. 
%% In the case of the DMRG and CP-AFQMC calculations, we also studied cluster sizes commensurate with multiple wavelengths to study competition between different wavelengths within the same finite system, or fit the stripe along the periodic width rather
%% than length of the cluster, in order to obtain multiple estimates of the energy of the same wavelength stripe.
%% We used DMRG 
%% clusters which could fit commensurates stripes with wavelengths 5-9, while the AFQMC
%% clusters could fit stripes with wavelengths 5-12.
Figure~\ref{fig:energy_wavelength} shows the energy per site of the stripe versus its wavelength $\lambda$ for the multiple
methods. Earlier DMRG calculations on the Hubbard model had focused on $\lambda=4$ (half-filled stripes) which are seen in HTSC's~\cite{white1998density,white2003stripes},
  but we now observe that these are relatively high in energy. A striking feature is that
for $\lambda= 5-8$ the energies are nearly degenerate. This is clearly seen in the DMET data 
where stripes of all wavelengths can be stabilized, as well as from the averaged energy of the methods
between $\lambda=5-8$ (stars in Fig.~\ref{fig:energy_wavelength}). 
The energy difference between the $\lambda=5$ and $\lambda=8$ stripe in the different methods is estimated to be between $0.0005t$ (DMRG)--$0.0041t$ (iPEPS). 
This suggests that the magnetic domain walls can fluctuate freely, consistent with proposals for fluctuating stripes. 
In particular, the stripes may be distorted at a small cost over long length scales. 

Although the different wavelengths are nearly degenerate, there appears
to be a slight  minimum near wavelength $\lambda=8$ (a filled stripe) in all the methods.
Very recently, similar filled stripes have been reported as the ground state in part of the frustrated $t$-$J$ model
phase diagram~\cite{dodaro2016intertwined}.
 $\lambda=9$ appears significantly higher in energy in both  DMET and DMRG.
In the DMRG calculations, the $\lambda=9$ state was not even metastable as boundary 
conditions and initial states were varied, so the high-energy state shown was forced with a static potential.
The AFQMC results show a much weaker dependence on wavelength for longer wavelengths, 
for example the $\lambda=8$ and $\lambda=10$
stripe energies per site appear to be within $0.0015t$. 
However, when a mixture of the $\lambda=8$ and $\lambda=10$ stripe states
is set up on a length 40 cluster that is commensurate with both, 
the state that survives is the $\lambda=8$ stripe, suggesting
a preference for this wavelength.
The increase in energy at wavelengths $\lambda > 8$ coincides with
unfavourable double occupancy of the stripe. This simple interpretation
is supported by a  mean-field (unrestricted Hartree-Fock (UHF)) calculation with an effective interaction $U/t=2.7$ chosen within
the self-consistent AFQMC procedure (Fig.~\ref{fig:energy_wavelength}, inset). The mean-field result shows a clear minimum at
a wavelength $8$ vertical stripe. (Note that this requires the use of an effective $U/t$; at the bare $U/t=8$, mean-field theory
would produce a diagonal stripe~\cite{Jie2013jpcm}). 
The correspondence between the energies and densities in the effective mean-field 
and correlated calculations suggests that  mean-field theory with a renormalized interaction may be surprisingly 
good at describing the energetics of stripes. 
However, mean-field theory appears to somewhat underestimate the degeneracy
of the stripes as a function of wavelength, particularly at shorter wavelengths.

\begin{figure}[htpb]
  \centering
  %\begin{subfigure}[b]{\columnwidth}
  %  \includegraphics[width=\textwidth]{w8stripe_dmet_new}
  %  % U8/DMET/8x2_pi    python ../../../plot_order.py DMET 8 2 numbered
  %  \caption{DMET}
  %\end{subfigure}
  %\begin{subfigure}[b]{\columnwidth}
  %  \includegraphics[width=\textwidth]{w8stripe_cpmc_new}
  %  % U8/CPMC/48x4    python ../../../plot_order.py CPMC 48 4 9,17,0,2,numbered,4
  %  \caption{AFQMC}
  %\end{subfigure}
  %\begin{subfigure}[b]{\columnwidth}
  %  \includegraphics[width=\textwidth]{w8stripe_ipeps_new}
  %  % U8/iPEPS/16x2    python ../../../plot_order.py iPEPS 16 2 numbered
  %  \caption{iPEPS}
  %\end{subfigure}
  %\begin{subfigure}[b]{\columnwidth}
  %  \includegraphics[width=\textwidth]{w8stripe_dmrg_new}
  %  % U8/DMRG/32x6_w8 python ../../../plot_order.py DMRG 32 6 8,16,1,3,numbered,4
  %  \caption{DMRG}
  %\end{subfigure}
  \includegraphics[width=\textwidth]{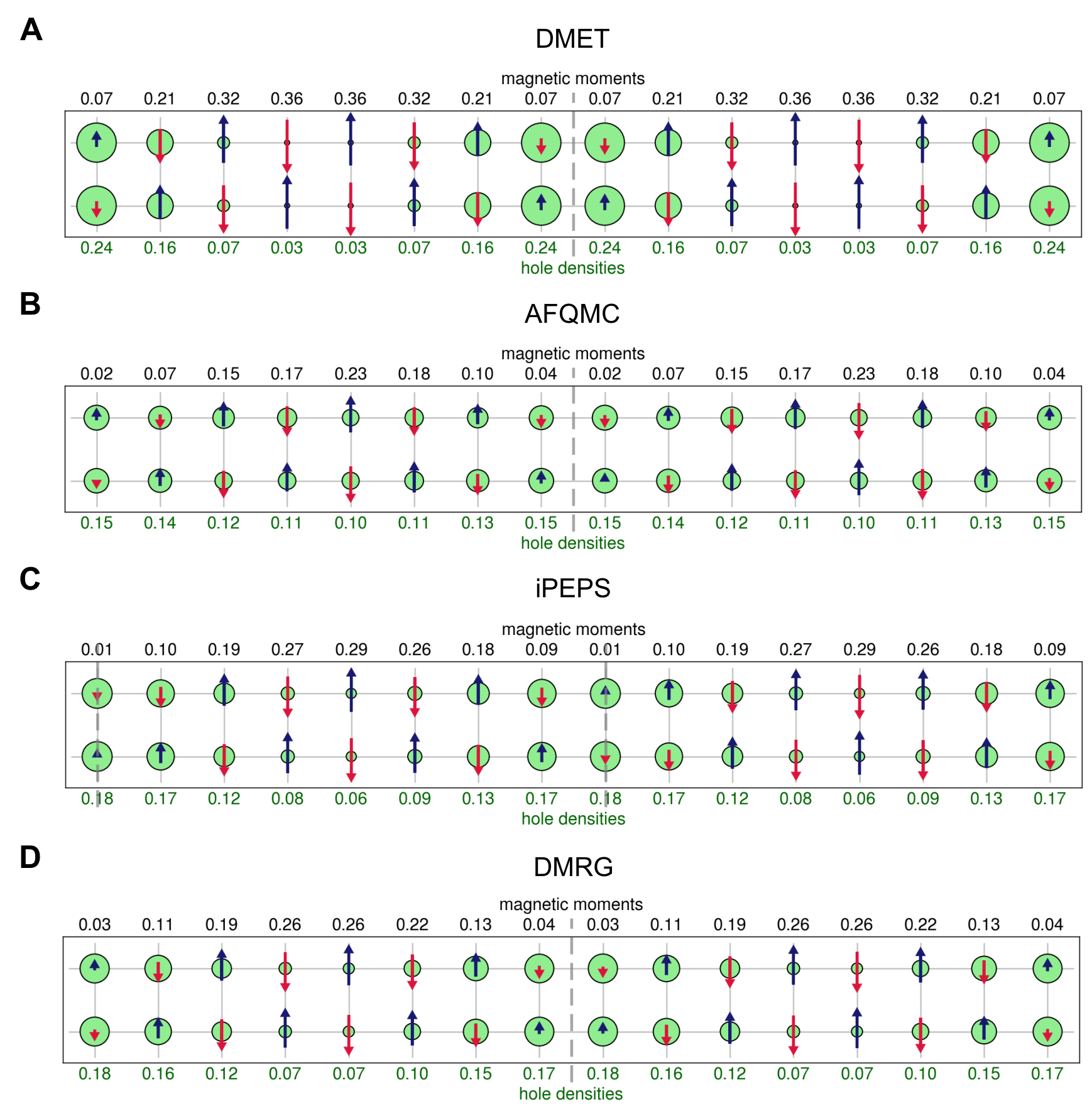}

  \caption{{\bf Charge and spin orders.} Shown are sketches of the charge and spin orders in the wavelength 8 stripes from {\bf(A)} DMET, {\bf(B)} AFQMC, {\bf(C)} iPEPS and {\bf(D)} DMRG. The local magnetic moments and hole densities are shown above and below   the order plots, respectively. (Circle radius is proportional to hole density, arrow height is proportional to spin density). The gray dashed lines represent the positions of maximum hole density and the magnetic domain wall. For more details, see \cite{supplementary}.}
  \label{fig:stripes}
\end{figure}

%% Our results indicate that the optimal stripe has wavelength
%%  $> 5$ (w5).
%% For wavelength $9$ DMET shows a significant increase in energy; 
%% further suggesting that this longer wavelength is higher in energy. The CP-AFQMC results with periodic boundary
%% conditions show a much weaker dependence on wavelength for long wavelengths, with some uncertainty
%% from the finite cluster size which was not extrapolated with cluster length, but the optimal wavelength
%% appears to be between wavelength $6$ and $12$.
%% DMET shows a significant
%% Wavelength 9 stripes could not be stabilized in the DMRG calculations
%% We also show the result for mean-field stripes (from Hartree-Fock theory) with an effective $U=2.8$ derived by
%% choosing an interaction that self-consistently matches the CP-AFQMC and Hartree-Fock densities through the
%% procedure described in Ref.~\cite{selfconsafqmc}. 

%% The DMRG and AFQMC data could be stabilized for stripes of wavelength 5 1/3, 7, and 8, while the iPEPS data shows a
%% slightly larger energy fluctuation between stripes of 5, 7 and wavelength 8. But in any case, the total energy scale is on the order of 1-2 millit.
%% As indicated in Fig.~\ref
%% The lowest energy is found for stripe 8 wavelengths. Note that although the charge order is for wavelength 8
%% the spin order is at wavelength 16. The wavelength 8 stripes are fully hole doped; earlier half-doped stripes correspond to using unit cells
%% with [something about filling here].

The vertical stripe order for the $\lambda= 8$ stripe from the different methods is depicted in Fig.~\ref{fig:stripes}. 
%% This is the lowest energy, or degenerate with the lowest energy stripe, to within statistical
%% and extrapolation error, in all the methods. 
We show the full period (16) for the spin modulation. The stripe is a  bond-centered stripe in the AFQMC, DMRG, and DMET calculations.
In the iPEPS calculation, the stripe is nominally site-centered.
In all the calculations, the width of the hole domain wall spans several sites,
blurring the distinction between bond- and site-centered stripes, and we conclude that the energy difference between
the two is very small. 
There is substantial agreement in the order observed by the different numerical techniques, with only some
differences in the modulation of the hole and spin-densities. 
%% The maximum magnetic moment ranges from $0.21$ (AFQMC) to $0.36$ (DMET), while the largest hole density (on the
%% stripe) ranges from $0.15$ (AFQMC) to $0.24$ (DMET). 

Note that for even wavelength stripes, the spin wavelength must be twice that of the charge modulation in order to 
accommodate the stripe as well as the antiferromagnetic order. At odd wavelengths,  site-centered stripes appear in all the calculations, and here charge and spin order can have the same wavelength. (This odd-even alternation
does not affect the peaks of the structure factor near $(\pi,\pi)$, see \cite{supplementary}).
%% {\color{red} It is possible to find site-centered stripes with
%% even wavelength (see e.g. Ref.~\cite{boxiao}), but these lie at considerably
%% higher energy ($>0.01t$ per site) than the bond-centered stripes. }
%% even wavelength str

\begin{figure}[htpb]
  \centering
  %\begin{subfigure}[b]{0.4\columnwidth}
  %  \includegraphics[width=\textwidth]{iPEPS_w5}
  %\caption{iPEPS $\lambda=5$}
  %\end{subfigure}
  %\begin{subfigure}[b]{0.55\columnwidth}
  %  \includegraphics[width=\textwidth]{iPEPS_w7}
  %\caption{iPEPS $\lambda=7$}
  %\end{subfigure}
  %\begin{subfigure}[b]{0.4\columnwidth}
  %  \includegraphics[width=\textwidth]{DMET_w5_sc}
  %  \caption{DMET metastable $\lambda=5$}
  %\end{subfigure}
  %\begin{subfigure}[b]{0.55\columnwidth}
  %  \includegraphics[width=\textwidth]{delta_replot}
  %  \caption{DMRG pairing order parameters.}
  %\end{subfigure}
  \includegraphics[width=\textwidth]{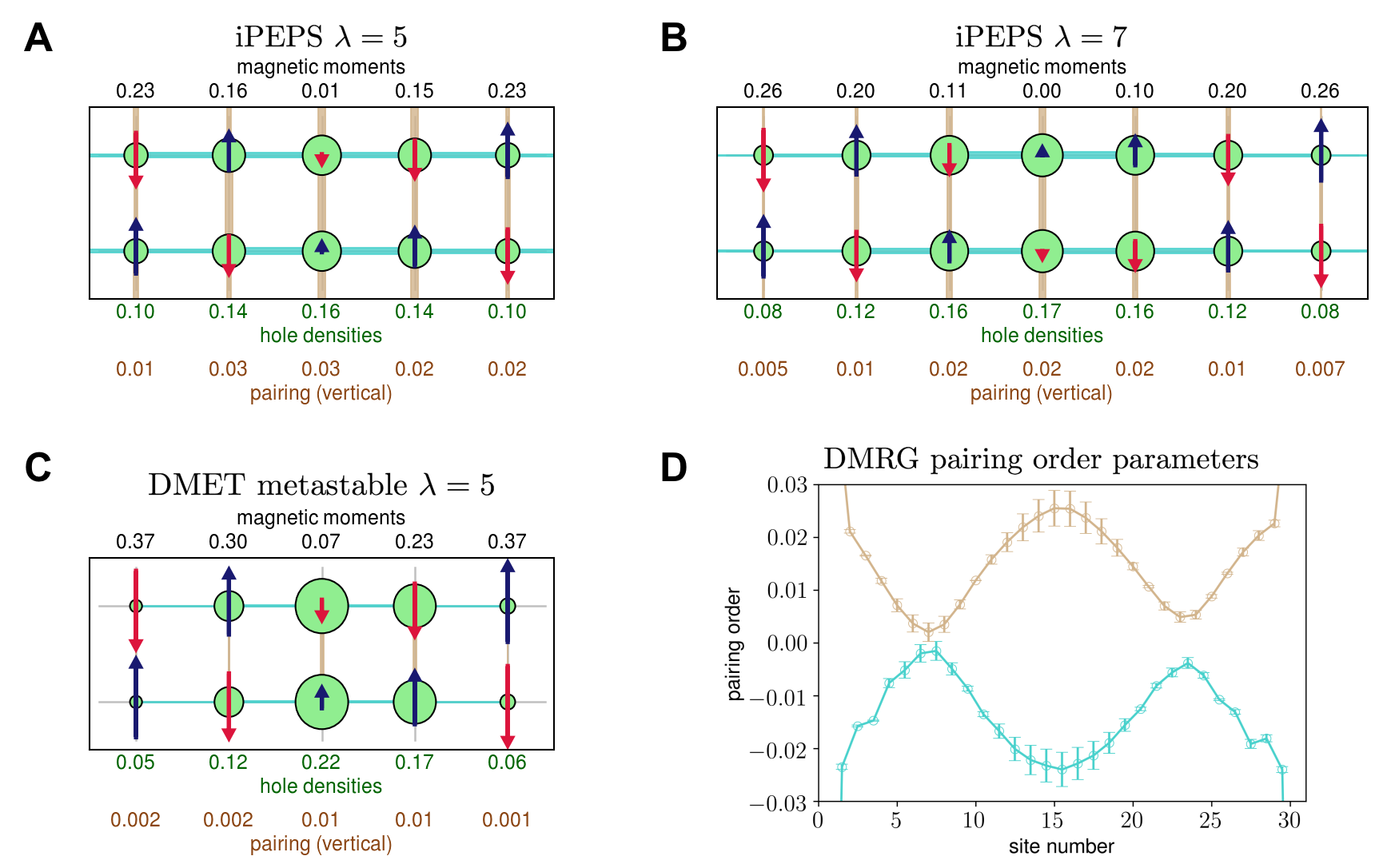}
  \caption{{\bf $d$-wave pairing.} Shown are metastable stripe states with $d$-wave pairing from iPEPS, DMET, and DMRG. {\bf(A), (B)} iPEPS stripes with $\lambda=5$ and $\lambda=7$. 
  {\bf(C)} DMET metastable $\lambda=5$ stripe with pairing. (Circle radius is proportional to hole density, arrow height is proportional to spin density, 
bond width is proportional to pairing density). {\bf(D)} DMRG pairing order parameters on a 32$\times$4 cylinder.
    The positive values are from the vertical bonds and the negative values from the horizontal bonds. $x$ axis is site number along the long-axis of the cylinder. For details, see \cite{supplementary}.}
  \label{fig:sc_stripes}
\end{figure}

\noindent {\bf Pairing order, fluctuations, and superconductivity}. A key question is whether pairing order coexists with stripe order. 
Previous work on the $t$-$J$ model with iPEPS found co-existing $d$-wave order for partially filled ($\lambda < 8$) stripes.
%% Filled stripes, i.e. one hole per lattice spacing along the stripe for vertical stripes, 
%% as observed in mean-field calculations 
%% It is usually argued that filled stripes, i.e. one hole per lattice spacing along the stripe for vertical stripes, as
%% or At the mean-field level stripes are always filled, i.e. one hole per lattice spacing along the stripe for vertical stripes, and 
%% do not exhibit pairing, in fact they are insulating. 
We did not find $d$-wave order in the Hubbard $\lambda=8$ stripe with any technique. 
However, $d$-wave order can be found at other wavelengths. For example, for $\lambda=5$, $\lambda=7$ stripes,
iPEPS produces $d$-wave order along the bonds (see Fig.~\ref{fig:sc_stripes}) with a maximum $d$-wave expectation value of $0.026$ and $0.021$, respectively. DMRG calculations with pinning pairing fields on the boundary for  a $32 \times 4$ cylinder also find $d$-wave order, with a maximum $d$-wave order of $0.025$, consistent with the iPEPS results. In the DMET calculations, the lowest energy $\lambda=5$ stripe
has no $d$-wave order, however, at slightly higher energy ($\sim 0.003t$) a $\lambda=5$ state similar to the iPEPS stripe can be
found with co-existing $d$-wave order, but with a substantially smaller maximum order parameter of $0.01$. Overall
our results support the coexistence of modulated $d$-wave order with the striped state, although the strength
of pairing is dependent on the stripe wavelength and filling. The pairing modulation we find  (Fig.~\ref{fig:sc_stripes})
is in-phase between cells. Other kinds of pairing inhomogeneities, such as pair density waves,
have also been discussed in the literature~\cite{Fradkin2015}. 

%% , but earlier $t$-$J$ simulations indicate that the similar anti-phase order
%% is almost degenerate on a scale of $0.001t$, {\color{red} consistent with a pair density wave}~\cite{Corboz2014,lee2014amperean}.

It has long been argued that fluctuating stripes could promote superconductivity~\cite{emery1997spin,kivelson1998electronic,zaanen2001}.
Our work provides some support for this picture, as there is a low-energy scale associated with
the deformation of stripe wavelength, and we also find coupling between the wavelength and the pairing channel.
We can imagine fluctuations in wavelength both at low temperatures, as well as in the ground-state. In the latter case,
this could lead to a stripe liquid ground-state rather than a stripe crystal. Our calculations are consistent with both possibilities.

\begin{figure}[htpb]
  \centering
  %\begin{subfigure}[b]{0.45\columnwidth}
  %\includegraphics[width=\columnwidth]{U6}
  %\caption{$U/t=6$.}
  %\label{fig:U6}
  %\end{subfigure}
  %\begin{subfigure}[b]{0.45\columnwidth}
  %\includegraphics[width=\columnwidth]{U12}
  %\caption{$U/t=12$.}
  %\label{fig:U12}
  %\end{subfigure}
  \includegraphics[width=\textwidth]{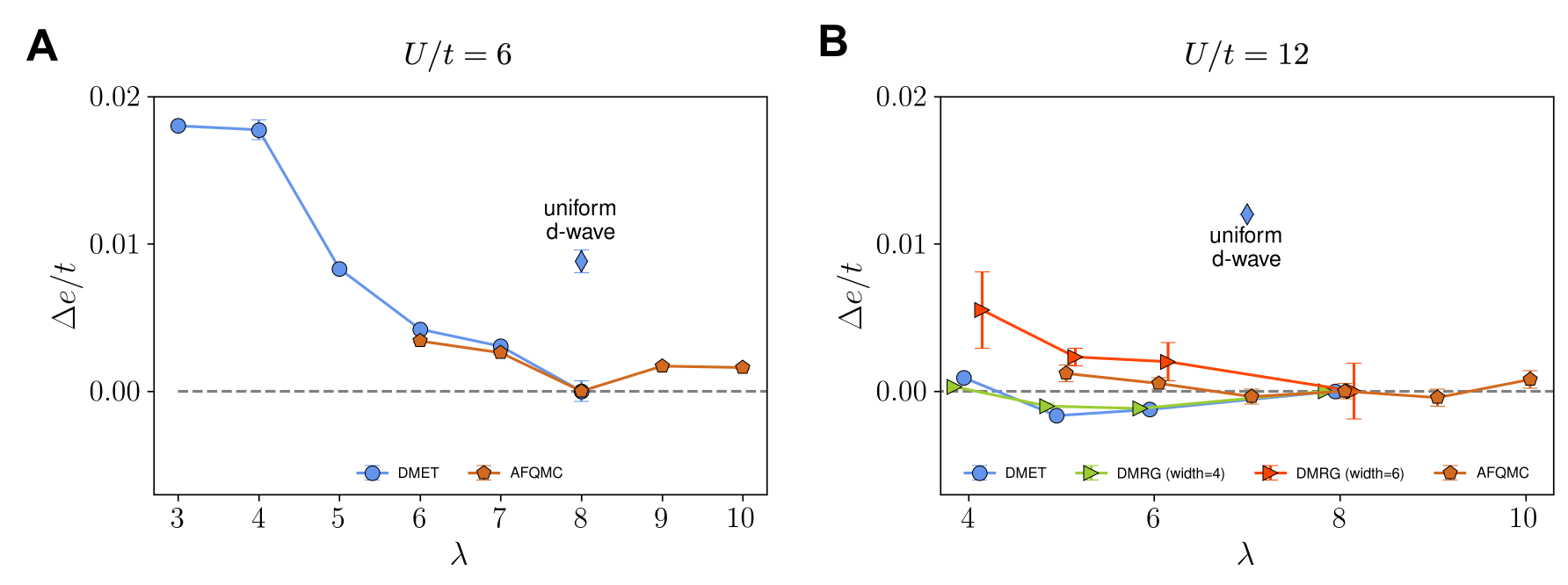}
  \caption{{\bf Varying the interaction strength.} {Relative energies of stripe states (vs. wavelength) and the uniform d-wave state at $1/8$ doping for {\bf(A)} weaker and {\bf(B)} stronger couplings. For details see \cite{supplementary}.}}
  \label{fig:coupling}
\end{figure}

\noindent {\bf Varying the coupling}. We may also ask whether the $U/t=8$, $1/8$ doping point 
is an anomalous point in the Hubbard phase diagram, and, if, for example, moving away from this
point would cause the unusual stripe compressibility (with respect to wavelength at fixed doping) to be lost. 
%% Other studies by this group and others indicate that competing orders are generally expected 
%% coupling produces an especially
%% degenerate 
%% A natural question to ask is whether or not the 1/8 doping point is an especially difficult point in the Hubbard phase diagram, for example
%% at the boundary between many different orders, and whether or not simply moving away from this point would alleviate the extreme degeneracies that we see.
{In Fig. \ref{fig:coupling} we show
the energies of various striped states and the uniform state at $U/t=6$ and $U/t=12$, $1/8$ doping, computed using AFQMC, DMET and DMRG.
At both couplings, the stripes around wavelength 8 are nearly degenerate, with the degeneracy increasing as the
coupling increases. At $U/t=6$, we find the ground state is a filled stripe state with wavelength $\lambda = 8$, with a larger energy
stabilization than at $U/t=8$. The trend is consistent with the state observed at $U/t=4$ with a more sinusoidal spin-density wave, more delocalized holes, and a more pronounced
minimum wavelength \cite{chang2010spin}. At $U/t=12$, we find a filled stripe with AFQMC and
DMRG (width 6), but DMET and DMRG on a narrower cylinder (width 4) find $\lambda= 5-6$. The similarity of the DMET and DMRG (width 4) data
suggests that the shorter wavelength is associated with a finite width effect.
We note that 2/3 filled stripes consistent with $\lambda= 5-6$ were also seen in
earlier DMRG studies on width 6 cylinders~\cite{hager2005stripe}, but a more detailed analysis
shows that the filled stripe becomes favoured when extrapolated to infinite bond dimension~\cite{supplementary}.
%over the 2/3 filled stripe at larger bond dimensions and cylinder lengths than those studied in Ref.~\cite{hager2005stripe}.
%% results, we find the filled
%% stripe has higher entanglement, and is only slightly favored as bond dimension $M$ goes beyond what was previously used (M=8000), and through extrapolation to
%% infinite bond dimension. 
Thus, we conclude that the ground state at $U/t=12$ is also the $\lambda=8$ stripe, although stripes of other wavelengths become
even more competitive than at $U/t=8$.
Overall, the similarity in the physics over a wide range of $U/t$ indicates that striped orders with low energy fluctuations of domain walls remain a robust feature in the moderate to strongly coupled underdoped region. 
}
%While it is costly to reproduce  the entire set of calculations across the phase diagram, in Fig. \ref{fig:U6} we show 
%the energies of various striped states and the uniform state at $U=6$, $1/8$ doping, computed using AFQMC and DMET.
%At this coupling, we similarly find that the ground state is a filled striped state with wavelength $\lambda \approx 8$. Further,
%the stripes around wavelength 8 have  similar energies, although the stripe order appears slightly less compressible than at $U=8$. 
%The trend is consistent with the state at $U=4$ of a more sinusoidal spin-density wave, more delocalized holes, and a more pronounced
%minimum wavelength \cite{chang2010spin}.
%Overall, the similarity between $U=6$ and $U=8$ physics suggests that striped orders with low energy fluctuations of domain walls will remain a robust feature of the physics in the moderate to strongly coupled underdoped region. 

\noindent {\bf Connection to stripe order in HTSC's}. 
In HTSC's the accepted stripe wavelength at 1/8 doping (e.g. in LaSrCuO) is $\lambda \approx 4.3$ (close to half-filled) ~\cite{tranquada1995evidence}.
%% The earliest experimental evidence for static stripes 
%% appeared in LaSrCuO at 1/8 doping, where half-filled stripes ($\lambda=4$) were observed Subsequent experiments on a variety of cuprates
%% at different dopings have similarly found charge ordering. 
{However, we find that the $\lambda=4$ stripe is
not favoured in the 2D Hubbard model for the coupling range ($U/t=6-12$) normally considered most relevant to cuprate physics}.
%This implies that physics beyond the Hubbard model is at play in the charge-ordering of real materials.
{This implies that the detailed charge-ordering of real materials arises from even stronger coupling or, more likely, quantitative corrections beyond the simple Hubbard model.}
{With respect to the latter,} one possibility is long-range hopping (such as a next-nearest neighbour hopping) which has been seen to change the preferred stripe wavelength
  in the frustrated $t$-$J$ model~\cite{dodaro2016intertwined}.
  Another possibility is the long-range Coulomb repulsion. Long-range repulsion
  can play a dual role, in both driving charge inhomogeneity, as well as smoothing it out.
  In the Hubbard model, where stripes naturally form, the
  latter property can help drive the ground state towards shorter stripe wavelengths.
  We have estimated the effect of the long-range interactions on the stripe energetics by computing
  the Coulomb energy of the charge distributions  in Fig. \ref{fig:stripes}. We use a dielectric constant of
  15.5 (in the range proposed for the cuprate plane~\cite{schuttler2001screening}). 
This gives a contribution favouring the shorter wavelength stripes that is $\sim O(0.01t)$
per site for the $\lambda=4$ versus $\lambda=8$ stripe~\cite{supplementary}. Although this is only an order of magnitude estimate, 
it is on the same energy scale as the stripe energetics in Fig.~\ref{fig:relative_energies}, and thus provides
a plausible competing mechanism for detailed stripe physics in real materials.

%% evidence for stripes in  appeared several years later, the stripes were half-filled~\cite{tranquada}.%[ref JM Tranquada, BJ Sternlieb, JD Axe, Y Nakamura, S Uchida Nature 375 (6532), 561-563.]

\section*{Conclusions}  In this work we have employed state-of-the-art numerical methods 
 to determine the ground state of the 1/8 doping point of the 2D Hubbard model at moderate to strong coupling. Through careful convergence of all the methods, and exhaustive cross-checks and validations, we are able to eliminate several of the competing orders that have been proposed for the underdoped region in favour of a vertical striped order with wavelength near $\lambda \approx 8$. The striped order displays a remarkably low energy scale associated with changing its wavelength, which
 implies strong fluctuations either at low temperature or in the ground-state itself. This low energy scale can roughly be accounted
 for at the mean-field level with
a strongly renormalized $U$. We find co-existing pairing order with a strength dependent on the stripe wavelength,
indicating a coupling of stripe fluctuations to superconductivity. The stripe degeneracy is robust as the coupling strength is varied.

%, and we find that the relative energetics can be significantly shifted through a long-range Coulomb term.

It has long been a goal of numerical simulations to provide definitive solutions of microscopic models. Our work demonstrates that even
in one of the most difficult condensed matter models, such unambiguous simulations are now  possible. In so far as the 2D Hubbard model
is a realistic model of high-temperature superconductivity, the stripe physics observed here provides a firm basis for understanding
the  diversity of inhomogeneous orders seen in the materials, as well as a numerical foundation for the theory
of fluctuations and its connections to superconductivity. However, our work also enables
us to see the limitations of the Hubbard model in understanding real HTSC's. 
Unlike the stripes at this doping point in real materials, we find filled stripes rather than near half-filled stripes. 
%% \red{While it might be true that the Hubbard model has all the relevant cuprate 
%% states as low energy possibilities, it seems too much to expect out of one oversimplified model for it to always find the correct ordering of these states.
\red{Given the very small energy scales involved, terms beyond the Hubbard model, such as long-range Coulomb interactions, will likely play a role in 
the detailed energetics of stripe fillings.} %Thus, it is necessary for numerical studies to investigate the impact of such terms.}
%The physics giving rise to half-filled stripes \delete{must}\red{may} therefore arise from
%terms beyond the Hubbard model, such as long-range Coulomb interactions.
%These will give
%contributions larger than the remaining inaccuracies in our simulations.
The work we have presented provides an optimistic perspective that achieving a comprehensive numerical characterization of
more detailed models of the HTSC's will also be within reach.

\nocite{schuttler2001screening,arrigoni2002stripes,qin2016coupling,pre_trial_wf,qin_2016_benchmark,LeBlanc2015,Motruk2016,verstraete2004,Verstraete08,nishino01,nishio2004,eisert2010,corboz16,corboz2010,phien15,jiang2008,corboz2011,nishino1996,orus2009-1,singh2010,bauer2011,Corboz09_fmera,knizia2012density,zheng2016,zheng2016cluster}

 \bibliographystyle{Science}

 \bibliography{ref,refs_additional,refs_ipeps,stripes}

\begin{thebibliography}{10}
\bibitem{pre_trial_wf}
We construct the trial wave-function with desired wave-length by solving the
  non-interacting Hamiltonian with pinning fields in the whole system with the
  same structure.
\end{thebibliography}

\begin{scilastnote}
\item[] Work performed by B.-X. Zheng, C.-M. Chung,
  M.-P. Qin, H. Shi, S. R. White, 
  S. Zhang, and G. K.-L. Chan was supported by the Simons Foundation through the Simons Collaboration
  on the Many-Electron Problem. S. R. White acknowledges support from the US NSF (DMR-1505406).
  S. Zhang and H. Shi acknowledge support from the US NSF (DMR-1409510). M. Qin was also supported by the US DOE (DE-SC0008627).
  G. K.-L. Chan acknowledges support from a Simons Investigatorship and the US DOE (DE-SC0008624).
  DMET calculations were carried out at the National Energy Research Scientific Computing Center, a US DOE Office of Science User Facility   supported by DE-AC02-05CH11231.   AFQMC calculations were carried out at the Extreme Science and
  Engineering Discovery Environment (XSEDE), 
  supported by the US NSF Grant No. ACI-1053575, at the OLCF at Oak Ridge National Lab, and the computational facilities at the
  College of William and Mary. P. Corboz was supported by the European Research Council (ERC) under the
  European Union's Horizon 2020 research and innovation programme (grant No 
  677061).
  G. Ehlers and R. M. Noack acknowledge support
  from the Deutsche Forschungsgemeinschaft (DFG) through grant
  no.\  NO 314/5-1 in Research Unit FOR 1807. Data used in this work is in the Supplementary Information and online at \url{github.com/zhengbx/stripe_data}.
  The DMET code is available online at \url{bitbucket.org/zhengbx/libdmet}. Other computer code is available from authors upon request: for AFQMC code, contact
  S. Zhang; for real-space DMRG code, contact S. R. White; contact R. M. Noack for hybrid DMRG code; and for iPEPS code, contact P. Corboz.
\end{scilastnote}

 \bigskip

 Supplementary information Sections S1-S9. Detailed description
 of all methods, data, and analysis. Supplementary Tables S1-S10, Figure S1-S41. References 51-69.

\end{document}